\documentclass[prd,aps,showkeys,superscriptaddress,two column]{revtex4-1}

\usepackage{amsmath}
\usepackage{amssymb}
\usepackage{amsthm}
\usepackage{mathrsfs}
\usepackage{graphicx}
\usepackage{fancyhdr}
\usepackage{array}
\usepackage{simplewick}
\usepackage{latexsym}
\usepackage[all]{xy}
\usepackage{eufrak}
\usepackage{euscript}
\usepackage{enumerate}
\usepackage{dsfont}
\usepackage{slashed}
\usepackage{hyperref}

\newcommand{\be}{\begin{equation}}
\newcommand{\ee}{\end{equation}}
\newcommand{\bs}{\begin{split}}
\newcommand{\es}{\end{split}}

\def \v {\vec}

\begin{document}



\title{Effective confinement theory from Abelian variables in $SU(3)$ gauge theory}

\author{L. S. Grigorio}
\email{leogrigorio@if.ufrj.br}
\affiliation{Instituto de F\'\i sica, Universidade Federal do Rio de Janeiro,
\\21941-972, Rio de Janeiro, Brazil}
\affiliation{Centro Federal de Educa\c{c}\~ao Tecnol\'ogica Celso Suckow da Fonseca,
\\28635-000, Nova Friburgo, Brazil}

\author{M. S. Guimaraes}
\email{msguimaraes@uerj.br}
\affiliation{Departamento de F\'\i sica Te\'orica, Instituto de F\'\i sica, UERJ - Universidade do
Estado do Rio de Janeiro, Rua S\~ ao Francisco Xavier 524, 20550-013 Maracan\~ a, Rio de Janeiro,
Brazil}

\author{R. Rougemont}
\email{romulo@if.ufrj.br}
\affiliation{Instituto de F\'\i sica, Universidade Federal do Rio de Janeiro,
\\21941-972, Rio de Janeiro, Brazil}

\author{C. Wotzasek}
\email{clovis@if.ufrj.br}
\affiliation{Instituto de F\'\i sica, Universidade Federal do Rio de Janeiro,
\\21941-972, Rio de Janeiro, Brazil}

\date{\today}

\begin{abstract}
In this Letter we use the Julia-Toulouse approach for condensation of defects in order to obtain an effective confinement theory for external chromoelectric probe charges in $SU(3)$ gauge theory in the regime with condensed chromomagnetic monopoles. We use the Cho decomposition of the non-Abelian connection in order to reveal the Abelian sector of the non-Abelian gauge theory and the associated topological defects (monopoles) without resorting to any gauge fixing procedure. Using only the Abelian sector of the theory, we construct a hydrodynamic effective theory for the regime with condensed defects in such a way that it is compatible with the Elitzur's theorem. The resulting effective theory describes the interaction between external chromoelectric probe charges displaying a short-range Yukawa interaction plus a linear confining term that governs the long distance physics.
\end{abstract}




\maketitle


\section{Introduction}
\label{sec:introduction}

There is a very popular proposal regarding the issue of static quark confinement that conjectures that the QCD vacuum should contain a condensate of chromomagnetic monopoles, constituting a dual superconductor which should generate an asymptotic linear chromoelectric confining potential through the dual Meissner effect \cite{conf}. Since monopoles are absent in pure Yang-Mills theories as classical solutions with finite energy, the dual superconductor scenario of color confinement is usually approached by using an Abelian projection \cite{abelproj}, which corresponds to some partial \emph{gauge fixing condition} implementing the following \emph{explicit breaking} pattern: $SU(N) \rightarrow U(1)^{N-1}$, followed by the discarding of the off-diagonal sector of the theory. It can be shown that under an Abelian projection the Abelian (diagonal) sector of the gluon field behaves like if chromomagnetic monopoles (whose chromomagnetic charges are defined with respect to the residual unfixed maximal Abelian subgroup $U(1)^{N-1}$) were sitting in points of the space where the Abelian projection becomes singular. It is then expected that if somehow these chromomagnetic defects proliferate (condense), the chromoelectric sources become confined; in particular, for the $SU(2)$ case, the lattice data \cite{suzuki} show the very interesting result that if one fixes the maximal Abelian gauge (MAG) and further discards the off-diagonal sector of the theory, then the Abelian confining string tension $\sigma_{U(1)}$ obtained reproduces $\sim 92\%$ of the full string tension $\sigma_{SU(2)}$, a phenomenon that is known as the Abelian dominance (for the confining string tension). For a review, see \cite{ripka} (see also the discussions in \cite{shabanov}).

Despite the success of the MAG calculations, there is an apparent problem of gauge dependence in such reasoning, since in the Abelian Polyakov gauge the monopole condensation does not lead to static quark confinement \cite{chernodub}. Indeed, within the Abelian projection approach it could appear in principle that monopoles are merely a gauge artifact. Since no physical phenomenon can depend on an arbitrary gauge choice, there are certainly some missing pieces in this puzzle. In particular, it seems that an important step to be taken is to develop a gauge independent way of revealing the Abelian sector of the Yang-Mills theory and the associated topological structures (monopoles). Regarding this point, a very interesting proposal was made by Cho \cite{cho-su2,cho-su3}, consisting in a reparameterization of the Yang-Mills connection, called the Cho decomposition, which has the feature of explicitly exposing its Abelian component and the associated topological structures without resorting to any gauge fixing procedure.

In this Letter we make use of a generalization of the Julia-Toulouse Approach (JTA) for condensation of defects \cite{jt,qt}, as put forward recently in \cite{artigao}, in order to obtain, out of the Abelian sector of the $SU(3)$ Yang-Mills theory singled out in a gauge independent way via the Cho decomposition of the $SU(3)$ connection \cite{cho-su3}, an effective confinement theory for static chromoelectric probe charges due to the condensation of chromomagnetic monopoles.

The main idea in the JTA is to allow one to study gauge theories in the presence of defects that eventually condense. If we are not interested in the details of the condensation process we can ask ourselves if, having the knowledge of the model that describes the system before the condensation, we are able to determine the effective model describing the system in the condensed regime. The condensate of topological defects establishes a new medium in which the defects constitute a continuous distribution in space. The low energy excitations of such a medium represent the new degrees of freedom of the condensed regime. Julia and Toulouse \cite{jt} specified a prescription to identify these new degrees of freedom, knowing the model that describes the regime with diluted defects. This prescription does not deal with the dynamical reasons responsible for the condensation process: this is considered a separate issue, beyond the scope of the prescription that is concerned only with the properties of the new degrees of freedom once the condensation of topological defects has taken place. However, the work of Julia and Toulouse has taken place in the context of ordered systems in condensed matter and due to the possible non-linearity of the topological currents, the absence of relativistic symmetry and the need for the introduction of dissipative terms in this scenario, the construction of effective actions is a very complicated issue.

Latter, the JTA was extended by Quevedo and Trugenberger \cite{qt} who showed that in theories involving $p$-forms, which are very common in effective descriptions of string theories, these difficulties do not show up. They showed that in this context the prescription can be defined into a more precise form, which leads to the determination of the effective action describing the system in the condensed regime. They have also shown that this leads naturally to the interpretation of the Abelian Higgs Mechanism as dual to the JTA. Based mainly on the ideas developed in \cite{jt,qt}, we recently developed a general procedure \cite{artigao} to address the condensation of these defects, whose main features are a careful treatment of a local symmetry called as brane symmetry (which is independent of the usual gauge symmetry, as discussed by Kleinert in \cite{mvf}), which consists in the freedom of deforming the unphysical Dirac strings without any observable consequences, and the development of the JTA to be completely compatible with the Elitzur's theorem \cite{elitzur}.


In \cite{artigao}, we also have shown that the JTA may be realized both in the direct space of the potentials (as in the proposal of \cite{qt}) or in its dual space, which is the realization chosen in the present work, as we shall discuss in section \ref{sec:alba}. The resulting effective theory here obtained with the joint use of the Cho decomposition and the JTA is compatible with the Elitzur's theorem in the condensed regime and describes the interaction between external chromoelectric charges displaying a short-range Yukawa interaction plus a linear confining term that governs the long distance physics.

\section{$SU(2)$ Cho decomposition}
\label{sec:su2}

In this section we review some relevant points of the Cho decomposition of the $SU(2)$ connection. The starting point
is the introduction of a unitary color triplet and Lorentz scalar $\hat{n}$ and the definition of the so-called restricted connection $\hat{A}_\mu$ which leaves $\hat{n}$ invariant under parallel transport on the principal bundle \cite{cho-su2},
\begin{equation}
\hat{D}_\mu\hat{n}:=\partial_\mu \hat{n} + g\hat{A}_\mu\times\hat{n}\equiv \v{0} \Rightarrow
\hat{A}_\mu=A_\mu\hat{n}-\frac{1}{g}\hat{n}\times\partial_\mu\hat{n},
\label{eq:1}
\end{equation}
where $g$ is the Yang-Mills coupling constant.

Due to the fact that the space of connections is an affine space, a general $SU(2)$ connection $\v{A}_\mu$ can be obtained from the restricted connection $\hat{A}_\mu$ by adding a field $\v{X}_\mu$ that is orthogonal to $\hat{n}$ \cite{cho-su2}.
Thus, the general form of the $SU(2)$ Cho decomposition is given by:
\begin{align}
&\v{A}_\mu=\hat{A}_\mu+\v{X}_\mu=A_\mu\hat{n}-\frac{1}{g}\hat{n}\times\partial_\mu\hat{n}+\v{X}_\mu,\nonumber\\
&\hat{n}^2=1\,\,\mbox{and}\,\,\hat{n}\cdot\v{X}_\mu=0.
\label{eq:2}
\end{align}

From the infinitesimal $SU(2)$ gauge transformation defined by:
\begin{align}
&\delta\v{A}_\mu=\frac{1}{g} D_\mu\v{\omega}:=\frac{1}{g}(\partial_\mu\v{\omega}+g\v{A}_\mu\times
\v{\omega}),\nonumber\\
&\delta\hat{n}=-\v{\omega}\times\hat{n},
\label{eq:3}
\end{align}
it follows that:
\begin{align}
&\delta A_\mu=\frac{1}{g}\hat{n}\cdot\partial_\mu\v{\omega},\nonumber\\
&\delta\hat{A}_\mu=\frac{1}{g}\hat{D}_\mu\v{\omega},\nonumber\\
&\delta\v{X}_\mu=-\v{\omega}\times\v{X}_\mu.
\label{eq:4}
\end{align}

We see from (\ref{eq:4}) that $A_\mu$ transforms like an $U(1)$ connection, being the abelian component of the $SU(2)$ connection explicitly revealed by the Cho decomposition without any gauge fixing procedure. Thus, we say that the unitary triplet field
$\hat{n}$ selects the Abelian direction in the internal color space for each spacetime point. Furthermore, we also see from (\ref{eq:3}) and (\ref{eq:4}) that the restricted connection $\hat{A}_\mu$ transforms like the general $SU(2)$ connection $\v{A}_\mu$ since the restricted covariant derivative is expressed in the adjoint representation, like the general covariant derivative, in terms of the $SU(2)$ structure constants defining the cross product. Hence, the restricted connection is already an $SU(2)$ connection carrying all the gauge degrees of freedom (but not all the dynamical degrees of freedom) of the non-Abelian gauge theory, being $\v{X}_\mu$ a vector-colored source term called the valence potential which carries the remaining dynamical degrees of freedom of the theory. Notice also from (\ref{eq:4}) that $\v{X}_\mu$ transforms covariantly as a matter field in the adjoint representation.


Using the covariant gauge fixing condition for the restricted connection,
\begin{equation}
\partial_\mu \hat{A}^\mu=\v{0}\Rightarrow\partial_\mu A^\mu=0\,\,\mbox{and}\,\,\hat{n}\times\partial^2\hat{n}-g A_\mu\partial^\mu\hat{n}=\v{0},
\label{eq:5}
\end{equation}
we see that the 2 independent components of the field $\hat{n}$ are completely fixed by a gauge condition and hence it is not a dynamical variable of the theory. The two dynamical variables of the theory are the fields $A_\mu$ and $\v{X}_\mu$.

The curvature tensor $\v{F}_{\mu\nu}$ associated to the gauge connection $\v{A}_\mu$ is given by:
\begin{align}
\v{F}_{\mu\nu}&=\partial_\mu\v{A}_\nu-\partial_\nu\v{A}_\mu+g\v{A}_\mu\times\v{A}_\nu \nonumber\\
&=\hat{F}_{\mu\nu}+\hat{D}_\mu\v{X}_\nu-\hat{D}_\nu\v{X}_\mu+g\v{X}_\mu\times\v{X}_\nu,
\label{eq:6}
\end{align}
where the restricted curvature tensor $\hat{F}_{\mu\nu}$ is given by:
\begin{align}
&\hat{F}_{\mu\nu}=\partial_\mu\hat{A}_\nu-\partial_\nu\hat{A}_\mu+g\hat{A}_\mu\times\hat{A}_\nu=
(F_{\mu\nu}+H_{\mu\nu})\hat{n},\label{eq:7}\\
&F_{\mu\nu}=\partial_\mu A_\nu-\partial_\nu A_\mu,\label{eq:8}\\
&H_{\mu\nu}=-\frac{1}{g}\hat{n}\cdot(\partial_\mu\hat{n}\times\partial_\nu\hat{n}).
\label{eq:9}
\end{align}

The Lagrangian density of the theory reads:
\begin{align}
\mathcal{L}&=-\frac{1}{4}\v{F}^2_{\mu\nu}\nonumber\\
&=-\frac{1}{4}\hat{F}^2_{\mu\nu}-\frac{1}{4}(\hat{D}_\mu\v{X}_\nu-\hat{D}_\nu\v{X}_\mu)^2+\nonumber\\
&-\frac{g}{2}\hat{F}^{\mu\nu}\cdot(\v{X}_\mu\times\v{X}_\nu)-\frac{g^2}{4}(\v{X}_\mu\times\v{X}_\nu)^2.
\label{eq:10}
\end{align}

The Euler-Lagrange equations of motion for $A_\mu$ are given by:
\begin{equation}
\partial^\mu(F_{\mu\nu}+H_{\mu\nu}+X_{\mu\nu})=-g\hat{n}\cdot[\v{X}^{\mu}\times(\hat{D}_
\mu\v{X}_\nu-\hat{D}_\nu\v{X}_\mu)],
\label{eq:11}
\end{equation}
where we defined:
\begin{equation}
X_{\mu\nu}:=g\hat{n}\cdot[\v{X}_{\mu}\times\v{X}_\nu],
\label{eq:12}
\end{equation}
while the Euler-Lagrange equations of motion for $\v{X}_\mu$ read:
\begin{equation}
\hat{D}^\mu(\hat{D}_\mu\v{X}_\nu-\hat{D}_\nu\v{X}_\mu)=g(F_{\mu\nu}+H_{\mu\nu}+X_{\mu\nu})\hat{n}\times\v{X}^\mu.
\label{eq:13}
\end{equation}

Notice that the constraints of the theory imply that $\v{X}_\mu\perp\hat{n}$ and $\hat{D}_\mu\v{X}_\nu\perp\hat{n}$, such that $\v{X^\nu}\times(\hat{D}_\mu\v{X}_\nu-\hat{D}_\nu\v{X}_\mu)\parallel\hat{n}$, and hence $g\v{X^\nu}\times(\hat{D}_\mu\v{X}_\nu-\hat{D}_\nu\v{X}_\mu)=g(\hat{n}\cdot[\v{X^\nu}\times(\hat{D}_\mu
\v{X}_\nu-\hat{D}_\nu\v{X}_\mu)])\hat{n}$. Thus, combining the equations of motion for the fields $A_\mu$ and $\v{X}_\mu$, $\hat{n}$(\ref{eq:11})+(\ref{eq:13}), we get the usual Yang-Mills equations of motion:
\begin{equation}
D^\mu\v{F}_{\mu\nu}=\v{0}.
\label{eq:14}
\end{equation}


We can obtain an explicit form for the unitary triplet $\hat{n}$ through an Euler rotation of the internal global Cartesian basis $\{\hat{e}_a,\,a=1,2,3\}$. Parameterizing the elements of the adjoint representation by the three Euler angles, $S=\exp(-\gamma J_3)\exp(-\theta J_2)\exp(-\varphi J_3)\in SO(3)$, we can define the local internal basis by $\{\hat{n}_a:=S^{-1}\hat{e}_a,\,a=1,2,3\}$, where $\hat{n}:=\hat{n}_3=S^{-1}\hat{e}_3=(\sin(\theta)\cos(\varphi), \sin(\theta)\sin(\varphi),\cos(\theta))$. With this parameterization for $\hat{n}$ it is easy to show that the monopole curvature tensor given by (\ref{eq:9}) reproduces the magnetic field generated by an antimonopole at the origin:
\begin{align}
H_{\mu\nu}&=-\frac{1}{g}\sin(\theta)(\partial_\mu\theta\partial_\nu\varphi-\partial_\nu\theta\partial_\mu\varphi)\nonumber\\
&=-(\delta_{\mu\theta}\delta_{\nu\varphi}-\delta_{\nu\theta}\delta_{\mu\varphi})\frac{1}{gr^2},
\label{eq:15}
\end{align}
where in the last line we identified the internal and the physical polar and azimuthal angles and used spherical coordinates to write the components of the gradient operator: $\partial_0:=\partial_t\equiv\partial/\partial t$, $\partial_1:=\partial_r\equiv\partial/\partial r$, $\partial_2:=\partial_{\theta}\equiv(1/r)\partial/\partial\theta$ and $\partial_3:=\partial_{\varphi}\equiv(1/r\sin(\theta))\partial/\partial\varphi$. In fact, eq. (\ref{eq:15}) gives us the first homotopy class of the mapping $\pi_2(SU(2)/U(1)\simeq S^2)=\mathbb{Z}$ defined by $\hat{n}$ (the second homotopy group is associated with the 2-sphere $S^2_{phys}$ at the spatial infinity of the physical space for each fixed time). The complete set of homotopy classes of this mapping defining the chromomagnetic charges of the theory is obtained by making the substitution $\varphi\mapsto m\varphi,\,m\in\mathbb{Z}$ in the above parameterization for $\hat{n}$ \cite{cho-su2}, from which we obtain the chromomagnetic charge of an antimonopole in the $m$-th homotopy class of $\pi_2(S^2)$:
\begin{align}
\tilde{g}(m):=\oint_{S^2_{phys}}d\v{S}\cdot\v{H}_{(m)}&=\oint_{S^2_{phys}}dS^i\frac{1}{2}\epsilon_{0i\mu\nu}H^{\mu\nu}_{(m)}
\nonumber\\
&=\int_{0}^{\pi}r^2d\theta\sin(\theta)\int_{0}^{2\pi}d\varphi\left(-\frac{m}{gr^2}\right)\nonumber\\
&=-\frac{4\pi m}{g},\,m\in\mathbb{Z},
\label{eq:16}
\end{align}
which is the non-Abelian version of the Dirac quantization condition \cite{dirac,ripka}.

The so-called magnetic gauge \cite{cho-su2} is defined by fixing the local color vector field $\hat{n}$ in the $\hat{e}_3$-direction in the internal space by means of the $S$-rotation. In this gauge, the restricted curvature tensor is written as $\hat{F}_{\mu\nu}=(F_{\mu\nu}+H_{\mu\nu})\hat{e}_3$. Defining the so-called magnetic potential by the expression:
\begin{equation}
\tilde{C}_\mu:=\frac{1}{g}(\cos(\theta)\partial_\mu\varphi+\partial_\mu\gamma),
\label{eq:17}
\end{equation}
we see that we can rewrite the monopole curvature tensor $H_{\mu\nu}$ as \cite{shabanov}:
\begin{equation}
H_{\mu\nu}=\partial_\mu\tilde{C}_\nu-\partial_\nu\tilde{C}_\mu-\chi_{\mu\nu},
\label{eq:18}
\end{equation}
where:
\begin{equation}
\chi_{\mu\nu}:=\frac{1}{g}(\cos(\theta)[\partial_\mu,\partial_\nu]\varphi+[\partial_\mu,\partial_\nu]\gamma),
\label{eq:19}
\end{equation}
is the chromomagnetic Dirac string associated to the monopole \cite{dirac}. It is easy to see now that the restricted connection transforms like $\hat{A}_\mu\mapsto(A_\mu+\tilde{C}_\mu)\hat{e}_3$ under the gauge transformation that leads us to the magnetic gauge.

The magnetic potential $\tilde{C}_\mu$ describes the potential of a monopole, being singular over its associated Dirac string, as we can easily see following the example discussed in \cite{oxman} - if we consider $\gamma=-\varphi$, we have from (\ref{eq:17}) that:
\begin{equation}
\tilde{C}_\mu=\frac{1}{g}(\cos(\theta)-1)\partial_\mu\varphi=\frac{1}{g}\frac{(\cos(\theta)-1)}{r\sin(\theta)}
\delta_{\mu\varphi},
\label{eq:20}
\end{equation}
which is the monopole potential singular over the Dirac string arbitrarily placed (by the choice made for $\gamma$) in the negative $\hat{e}_3$-axis ($\theta=\pi$). Notice, however, that the monopole curvature tensor (\ref{eq:18}) is regular, as it should be.

If we now discard the valence potential $\v{X}_\mu$, that is, if we discard the off-diagonal sector of the theory, we end up with the restricted Lagrangian density defining the Abelian sector of the gauge theory:
\begin{align}
\mathcal{L}^{(R)}&=-\frac{1}{4}\hat{F}^2_{\mu\nu} \nonumber\\
&=-\frac{1}{4}[\partial_\mu(A_\nu+\tilde{C}_\nu)-\partial_\nu(A_\mu+\tilde{C}_\mu)-\chi_{\mu\nu}]^2.
\label{eq:21}
\end{align}
Notice that, being orthogonal to the internal unitary triplet $\hat{n}:=\hat{n}_3$, the valence potential $\vec{X}_\mu$ features 8 degrees of freedom. In fact, we can write $\vec{X}_\mu=X_\mu^1\hat{n}_1+X_\mu^2\hat{n}_2$, where each vector field $X_\mu^i,\,i=1,2$, has 4 degrees of freedom. The gauge transformation law (\ref{eq:4}) for the valence potential corresponds to transformations of the local internal base vectors $\hat{n}_i,\,i=1,2$, while the components $X_\mu^i,\,i=1,2$, of $\vec{X}_\mu$ do not change at all. Hence, the discarding of the off-diagonal degrees of freedom of the theory described by the vector fields $X_\mu^i,\,i=1,2$, which implies in the discarding of the valence potential $\vec{X}_\mu$, constitutes a color gauge invariant approximation which distinguishes our approach from the Abelian projection method.

If we further absorb the singular magnetic potential $\tilde{C}_\mu$ into the regular Abelian gluon field $A_\mu$ by making the shift $(A_\mu+\tilde{C}_\mu)\mapsto A_\mu$, we rewrite (\ref{eq:21}) in the following remarkable simple form:
\begin{align}
\mathcal{L}^{(R)}=-\frac{1}{4}(F_{\mu\nu}-\chi_{\mu\nu})^2,
\label{eq:22}
\end{align}
where now the field $A_\mu$ is singular over the chromomagnetic Dirac strings. Equation (\ref{eq:22}) describes the Maxwell
theory with the vector potential $A_\mu$ non-minimally coupled to monopoles. In \cite{jt-cho-su2} we added a minimal coupling of the gauge field with external chromoelectric probe charges to this Lagrangian density and applied a generalization \cite{mcsmon,dafdc,jt-cho-su2,artigao} of the Julia-Toulouse approach for condensation of defects, compatible with the Elitzur's theorem \cite{elitzur}, to obtain a confining low energy effective theory for these charges in the static limit assuming the establishment of a stable chromomagnetic condensate (in \cite{cho-sno} it is proposed a demonstration of the establishment of such a stable chromomagnetic condensation in the $SU(2)$ gauge theory). We shall take these steps in details in section \ref{sec:alba} for the $SU(3)$ case.

\section{$SU(N)$ and $SU(3)$ Cho decompositions}
\label{sec:su3}

The generalization of the $SU(2)$ Cho decomposition (\ref{eq:2}) for the $SU(N)$ case proposed by Shabanov is given by \cite{shabanov}:
\begin{align}
&\v{A}_\mu=\hat{A}_\mu+\v{X}_\mu=\sum_{(k)}\left[A_\mu^{(k)}\hat{n}^{(k)}-\frac{1}{g}\hat{n}^{(k)}\times
\partial_\mu\hat{n}^{(k)}\right]+\v{X}_\mu,\nonumber\\
&\hat{n}_{(i)}\cdot\hat{n}_{(k)}=\delta_{(i)(k)}\,\,\mbox{and}\,\,\hat{n}_{(k)}\cdot\v{X}_\mu=0,
\label{eq:23}
\end{align}
where the referred ``cross product" is defined by the $SU(N)$ structure constants and the indices between parentheses refer to the $(N-1)$ elements of the Cartan subalgebra of $SU(N)$.

Since our main objective in the present Letter is to investigate the contribution of the Abelian sector of the $SU(3)$ gauge theory to the issue of static chromoelectric confinement, from now on we focus only on the $SU(3)$ restricted connection, given by \cite{cho-su3}:
\begin{align}
&\hat{A}_\mu=\sum_{(k)=3,8}\left[A_\mu^{(k)}\hat{n}^{(k)}-\frac{1}{g}\hat{n}^{(k)}\times\partial_\mu\hat{n}^{(k)}\right],
\nonumber\\
&\hat{n}_{(3)}^2=\hat{n}_{(8)}^2=1\,\,\mbox{and}\,\,\hat{n}_{(3)}\cdot\hat{n}_{(8)}=0.
\label{eq:24}
\end{align}

The restricted connection (\ref{eq:24}) is the connection that leaves the unitary octets $\hat{n}^{(3)}$ and $\hat{n}^{(8)}$ invariant under parallel transport on the principal bundle. Actually, the magnetic condition $\hat{D}_\mu\hat{n}^{(3)}=\v{0}$  automatically generates the field $\hat{n}^{(8)}$ through the definition $\hat{n}^{(8)}:=\sqrt{3}\hat{n}^{(3)}\star
\hat{n}^{(3)}$, where the star product is defined by the symmetric $SU(3)$ $d$-constants. The chromomagnetic charges of the theory are labeled by two integers according to the mapping $\pi_2(SU(3)/U(1)\times U(1))=\mathbb{Z} \times\mathbb{Z}$ defined by $\hat{n}^{(3)}$, where $U(1)\times U(1)$ is the maximal Abelian subgroup of $SU(3)$ selected in each spacetime point in a gauge independent way by $\hat{n}^{(3)}$ and $\hat{n}^{(8)}$:
\begin{align}
\tilde{\v{g}}(N,N')=\tilde{g}\left(N-\frac{N'}{2},\sqrt{3}\frac{N'}{2}\right),\,\,N,N'\in\mathbb{Z},\,\,\tilde{g}=\frac{4\pi}{g}.
\label{eq:25}
\end{align}
Analogously to the discussion made around equation (\ref{eq:16}), the components $(3)$ and $(8)$ of $\tilde{\v{g}}(N,N')$ are defined as being the chromomagnetic fluxes of the types $(3)$ and $(8)$ through $S^2_{phys}$ by using explicit parameterizations for $\hat{n}^{(3)}$ and $\hat{n}^{(8)}$ obtained by means of a rotation of the internal global base elements $\hat{e}^{(3)}$ and $\hat{e}^{(8)}$ realized by an arbitrary element of the adjoint representation of $SU(3)$ parameterized by eight ``Euler angles". As we can see from (\ref{eq:25}), this is equivalent to take $\tilde{\v{g}}(N,N')$ as a linear combination with integer coefficients of the simple roots of the $SU(3)$ algebra. Also, as before, the magnetic potentials $\tilde{C}_\mu^{(3)}$ and $\tilde{C}_\mu^{(8)}$ are revealed in the restricted connection by fixing the magnetic gauge which is defined by an internal rotation that sends the local internal vectors $\hat{n}^{(3)}$ and $\hat{n}^{(8)}$ to the $\hat{e}^{(3)}$ and $\hat{e}^{(8)}$ directions, respectively (see \cite{cho-su3} and references therein for details).

The chromoelectric charge operator in the fundamental representation of $SU(3)$ is given by (see appendix D.2 of \cite{ripka}):
\begin{align}
\v{\mathcal{G}}:=g\v{\mathbb{T}}=g\left(\mathbb{T}_{(3)},\mathbb{T}_{(8)}\right),\,\,\mathbb{T}_{(k)}=
\frac{1}{2}\mathbb{\lambda}_{(k)},
\label{eq:26}
\end{align}
where $\mathbb{\lambda}_{(k)}$ are Gell-Mann matrices.

The restricted curvature tensor in the magnetic gauge is given by (absorbing, as before, the singular magnetic potentials $\tilde{C}_\mu^{(k)}$ into the regular Abelian gluon fields $A_\mu^{(k)}$):
\begin{align}
\hat{F}_{\mu\nu}=\sum_{(k)=3,8}(F_{\mu\nu}^{(k)}-\chi_{\mu\nu}^{(k)})\hat{e}^{(k)},
\label{eq:27}
\end{align}
where, as before, the chromomagnetic string terms $\chi_{\mu\nu}^{(k)}$ appear due to the angular (multivalued) nature of the components of the magnetic potentials $\tilde{C}_\mu^{(k)}$. The associated Lagrangian density is given by:
\begin{align}
\mathcal{L}^{(R)}=-\frac{1}{4}\sum_{(k)=3,8}(F_{\mu\nu}^{(k)}-\chi_{\mu\nu}^{(k)})^2.
\label{eq:28}
\end{align}

If we now minimally couple external chromoelectric probe charges to the restricted connection in the magnetic gauge, we get the following Lagrangian density:
\begin{align}
\bar{\mathcal{L}}^{(R)}=&\sum_{(k)=3,8}\left[-\frac{1}{4}(F_{\mu\nu}^{(k)}-\chi_{\mu\nu}^{(k)})^2-j_\mu^{(k)}A^\mu_{(k)}\right].
\label{eq:29}
\end{align}

We must now specify the chromoelectric charge structure of the probe current $\v{j}_\mu=(j_\mu^{(3)},j_\mu^{(8)})$. This can be done by comparison with the quarkionic current $\v{j}_\mu^\psi=\bar{\psi}\gamma_\mu\v{\mathcal{G}}\psi$ that couples minimally to the gauge connection $\hat{A}_\mu=(A_\mu^{(3)}+\tilde{C}_\mu^{(3)})\hat{e}^{(3)}+(A_\mu^{(8)}+ \tilde{C}_\mu^{(8)})\hat{e}^{(8)}$:
\begin{align}
\v{j}_\mu^\psi\cdot\hat{A}^\mu&=\bar{r}\gamma_\mu\left[\frac{g}{2}(A^\mu_{(3)}+\tilde{C}^\mu_{(3)})+
\frac{g}{2\sqrt{3}}(A^\mu_{(8)}+\tilde{C}^\mu_{(8)})\right]r+\nonumber\\
&+\bar{b}\gamma_\mu\left[-\frac{g}{2}(A^\mu_{(3)}+\tilde{C}^\mu_{(3)})+
\frac{g}{2\sqrt{3}}(A^\mu_{(8)}+\tilde{C}^\mu_{(8)})\right]b+\nonumber\\
&+\bar{y}\gamma_\mu\left[-\frac{g}{\sqrt{3}}(A^\mu_{(8)}+\tilde{C}^\mu_{(8)})\right]y,
\label{eq:30}
\end{align}
where $\bar{\psi}=(\bar{r},\bar{b},\bar{y})$ is the internal antitriplet, being $\bar{r}$, $\bar{b}$ and $\bar{y}$ the red, blue and yellow antiquark spinors, respectively. From (\ref{eq:30}) we see that the red, blue and yellow quarks have chromoelectric charges $\v{Q}=(Q_{(3)},Q_{(8)})$ given by $(\frac{g}{2},\frac{g}{2\sqrt{3}})$, $(-\frac{g}{2},\frac{g}{2\sqrt{3}})$ and $(0,-\frac{g}{\sqrt{3}})$, respectively (see also table (D.10) of \cite{ripka} and \cite{cho-su3}).

We now construct the chromoelectric probe current $\v{j}_\mu=(j_\mu^{(3)},j_\mu^{(8)})$ in such a way that it is compatible with the above charge structure:
\begin{align}
j_\mu^{(3)}&=\frac{g}{2}\delta_\mu(x;L_r)-\frac{g}{2}\delta_\mu(x;L_b)\nonumber\\
&=\frac{1}{2}\epsilon_{\mu\nu\alpha\beta}\partial^\nu\Lambda^{\alpha\beta}_{(3)},
\label{eq:31}
\end{align}
\begin{align}
\Lambda^{\alpha\beta}_{(3)}=\frac{g}{2}\tilde{\delta}^{\alpha\beta}(x;S_r)-\frac{g}{2}\tilde{\delta}^{\alpha\beta}(x;S_b),
\label{eq:32}
\end{align}
\begin{align}
j_\mu^{(8)}&=\frac{g}{2\sqrt{3}}\delta_\mu(x;L_r)+\frac{g}{2\sqrt{3}}\delta_\mu(x;L_b)
-\frac{g}{\sqrt{3}}\delta_\mu(x;L_y)\nonumber\\
&=\frac{1}{2}\epsilon_{\mu\nu\alpha\beta}\partial^\nu\Lambda^{\alpha\beta}_{(8)},
\label{eq:33}
\end{align}
\begin{align}
\Lambda^{\alpha\beta}_{(8)}=\frac{g}{2\sqrt{3}}\tilde{\delta}^{\alpha\beta}(x;S_r)+\frac{g}{2\sqrt{3}}
\tilde{\delta}^{\alpha\beta}(x;S_b)-\frac{g}{\sqrt{3}}\tilde{\delta}^{\alpha\beta}(x;S_y),
\label{eq:34}
\end{align}
where $\Lambda_{\mu\nu}^{(k)}$ are the chromoelectric Dirac string terms, being $L_r=\partial S_r$, $L_b=\partial S_b$ and $L_y=\partial S_y$ the worldlines of the red, blue and yellow probe charges, boundaries of the worldsurfaces $S_r$, $S_b$ and $S_y$ of the chromoelectric Dirac strings of the red, blue and yellow probe charges, respectively.

\section{The effective theory of color confinement}
\label{sec:alba}

If we now assume that in a certain regime of the theory it is established a stable chromomagnetic monopole condensate, we can ask ourselves what should be the form of the effective theory describing the low energy excitations of such a condensate.

One interesting approach to the general issue of the determination of the form of the effective field theory describing a regime with condensed defects, having previous knowledge of the form of the theory in the regime where these defects are diluted, was introduced by Julia and Toulouse within the context of ordered solid-state media \cite{jt} and further developed by Quevedo and Trugenberger within the relativistic field theory context \cite{qt}, constituting the so-called Julia-Toulouse approach (JTA) for condensation of defects. Another interesting approach to this same issue was developed by Banks, Myerson and Kogut within the context of relativistic lattice field theories \cite{banks} and also by Kleinert within the condensed matter context \cite{mvf}, constituting what we called in \cite{dafdc} the Abelian lattice based approach (ALBA). In a recent work \cite{artigao}, we unified and generalized these two approaches. By convention, we keep calling this generalized approach for the determination of the effective theory describing the regime with condensed defects, simply as JTA. 

As already mentioned, the JTA can be worked out either in the direct space of the potentials or in its dual space. Since we are interested in analyzing some of the consequences of a monopole condensation, it is interesting to go to the dual picture, where we avoid the problem of working with the vector potentials $A_\mu^{(k)}$ in a scenario where they are ill-defined in almost the whole spacetime due to the proliferarion of the chromomagnetic Dirac strings. Thus, our first step consists in dualizing the action associated to the restricted Lagrangian density (\ref{eq:29}) \cite{dafdc}:
\begin{align}
*\bar{S}^{(R)}=\int d^4x\sum_{(k)=3,8}\left[-\frac{1}{4}(\tilde{F}_{\mu\nu}^{(k)}-\Lambda_{\mu\nu}^{(k)})^2+ \tilde{j}_\mu^{(k)}\tilde{A}^\mu_{(k)}\right],
\label{eq:35}
\end{align}
where the couplings are inverted relatively to the ones present in (\ref{eq:29}) - here the dual vector potentials
$\tilde{A}_\mu^{(k)}$ couple minimally to the monopoles and non-minimally to the chromoelectric charges. Hence, in the dual picture the chromoelectric charges are seen as defects by the dual vector potentials, which are singular over the associated chromoelectric strings $\Lambda_{\mu\nu}^{(k)}$ given by (\ref{eq:32}) and (\ref{eq:34}).

Notice that the minimal coupling has support only over the chromomagnetic worldlines and, since the dual potentials are singular over the worldsurfaces of the chromoelectric Dirac strings (which we call as ``chromoelectric Dirac branes'', the term ``brane'' here meaning a generic hypersurface embedded in spacetime), the minimal coupling would be singular in events of the spacetime where the chromoelectric Dirac branes cross the chromomagnetic worldlines. Hence, in order to the action (\ref{eq:35}) to be regular everywhere in spacetime, \emph{the chromoelectric Dirac branes must not cross the chromomagnetic worldlines}, which is the dual version of the famous \emph{Dirac's veto} \cite{dirac}. Due to the Dirac's veto, the so-called \emph{chromoelectric Dirac brane symmetry corresponds to the freedom of moving the unphysical chromoelectric Dirac branes through the geometric place of the spacetime not occupied by the chromomagnetic monopoles}.

We must now specify the chromomagnetic charge structure that shall undergo a condensation process. Notice from (\ref{eq:25}) that there are three monopoles of minimal chromomagnetic charge (in modulus) in the non-trivial mapping $\pi_2(SU(3)/U(1)\times U(1))$, namely: $\tilde{\v{g}}_1:=\tilde{\v{g}}(1,0)=\frac{4\pi}{g}\v{\omega}_1$, $\tilde{\v{g}}_2:=\tilde{\v{g}}(-1,-1)=\frac{4\pi}{g}\v{\omega}_2$ and $\tilde{\v{g}}_3:=\tilde{\v{g}}(0,1)=\frac{4\pi}{g}\v{\omega}_3$, where $\v{\omega}_1=\left(1,0\right)$, $\v{\omega}_2=\left(-\frac{1}{2},-\frac{\sqrt{3}}{2}\right)$ and $\v{\omega}_3=\left(-\frac{1}{2},\frac{\sqrt{3}}{2}\right)$ are the positive roots of the $SU(3)$ algebra. The corresponding antimonopole charges (negative roots) are obtained by inverting the signs of $N$ and $N'$ in the previous configurations. It is energetically favorable that the lowest chromomagnetic charges (in modulus) allowed by the mapping $\pi_2(SU(3)/U(1)\times U(1))$ undergo a condensation process, thus, since we are going to deal with three condensing monopole currents of minimal chromomagnetic charge, we write:
\begin{align}
&\tilde{j}_\mu^i:=\tilde{g}\delta_\mu(x;\tilde{L}_i)=\frac{1}{2}\epsilon_{\mu\nu\alpha\beta}\partial^\nu
\chi^{\alpha\beta}_i,\label{eq:36}\\
&\chi^{\alpha\beta}_i=\tilde{g}\tilde{\delta}^{\alpha\beta}(x;\tilde{S}_i),\label{eq:37}
\end{align}
where $\tilde{L}_i=\partial\tilde{S}_i,\,i=1,2,3$ are the worldlines of the three monopoles associated to the roots $\v{\omega}_i$, the physical boundaries of the worldsurfaces $\tilde{S}_i$ of the unphysical chromomagnetic Dirac strings. The roots appear explicitly in the minimal coupling $\tilde{j}_\mu^i \tilde{A}^\mu_i$, where we defined $\tilde{A}^\mu_i:=\v{\omega}_i\cdot\tilde{\v{A}}_\mu=\v{\omega}_i\cdot(\tilde{A}_\mu^{(3)},\tilde{A}_\mu^{(8)})$.

In order to rewrite (\ref{eq:35}) in an explicitly Weyl-symmetric form (the Weyl symmetry corresponds to the invariance of the theory under permutations of the indices $i=1,2,3$ of the fundamental representation of $SU(3)$), we shall make use of the following identities:
\begin{align}
&\v{V}_\mu=(V_\mu^{(3)},V_\mu^{(8)})=\frac{2}{3}\sum_{i=1}^{3}\v{\omega}_i V_\mu^i,\nonumber\\
&V_\mu^i:=\v{\omega}_i\cdot\v{V}_\mu\Rightarrow\sum_{i=1}^{3} V_\mu^i=0\,\,\,\mbox{(constraint)};\nonumber\\
&\v{V}_\mu\cdot\v{R}^\mu=V_\mu^{(3)}R^\mu_{(3)}+V_\mu^{(8)}R^\mu_{(8)}=\frac{2}{3}\sum_{i=1}^{3} V_\mu^i R^\mu_i.
\label{eq:38}
\end{align}

Thus, the Weyl-symmetric representation of (\ref{eq:35}) with the monopole current given by (\ref{eq:36}) is:
\begin{align}
*\bar{S}^{(R)}=\int d^4x\sum_{i=1}^{3}\left[-\frac{1}{6}(\tilde{F}_{\mu\nu}^i-\Lambda_{\mu\nu}^i)^2+ \tilde{j}_\mu^i\tilde{A}^\mu_i\right],
\label{eq:39}
\end{align}
where the chromoelectric branes acquire the following remarkable symmetric form:
\begin{align}
\Lambda_{\mu\nu}^1&=\v{\omega}_1\cdot\v{\Lambda}_{\mu\nu}=\Lambda_{\mu\nu}^{(3)}\nonumber\\
&=\frac{g}{2}\tilde{\delta}_{\mu\nu}(x;S_r)-\frac{g}{2}\tilde{\delta}_{\mu\nu}(x;S_b),
\label{eq:40}
\end{align}
\begin{align}
\Lambda_{\mu\nu}^2&=\v{\omega}_2\cdot\v{\Lambda}_{\mu\nu}=-\frac{1}{2}\Lambda_{\mu\nu}^{(3)}-\frac{\sqrt{3}}{2}
\Lambda_{\mu\nu}^{(8)}\nonumber\\
&=-\frac{g}{2}\tilde{\delta}_{\mu\nu}(x;S_r)+\frac{g}{2}\tilde{\delta}_{\mu\nu}(x;S_y),
\label{eq:41}
\end{align}
\begin{align}
\Lambda_{\mu\nu}^3&=\v{\omega}_3\cdot\v{\Lambda}_{\mu\nu}=-\frac{1}{2}\Lambda_{\mu\nu}^{(3)}+\frac{\sqrt{3}}{2}
\Lambda_{\mu\nu}^{(8)}\nonumber\\
&=\frac{g}{2}\tilde{\delta}_{\mu\nu}(x;S_b)-\frac{g}{2}\tilde{\delta}_{\mu\nu}(x;S_y).
\label{eq:42}
\end{align}

In order to allow the monopoles to proliferate we must give dynamics to their associated chromomagnetic Dirac branes $\chi_{\mu\nu}^i$, since the proliferation of them is directly related to the proliferation of the monopoles and their worldlines. Thus, our second step consists in supplementing the dual action (\ref{eq:39}) with a kinetic term for the chromomagnetic Dirac branes of the form $-\frac{1}{2m^2}\tilde{j}_{\mu i}^2$, which is the term in a derivative expansion with the lowest order in derivatives of $\chi_{\mu\nu}^i$ (that is, the dominant contribution for the hydrodynamic limit of the theory) satisfying the relevant (Lorentz, gauge and brane) symmetries of the system. Such a contribution corresponds to an activation term for the chromomagnetic loops \cite{mvf}. Hence, the partition function associated to the extended dual action describing the regime with condensed chromomagnetic monopoles reads:
\begin{align}
&Z_c:=\prod_{i=1}^{3}\int\mathcal{D}\tilde{A}_\mu^i\,\delta[\partial_\mu\tilde{A}^\mu_i]
e^{i\int d^4x\left[-\frac{1}{6}(\tilde{F}_{\mu\nu}^i-\Lambda_{\mu\nu}^i)^2\right]}Z_c[\tilde{A}_\mu^i],
\label{eq:43}
\end{align}
where the Lorentz gauge was adopted for the dual potentials $\tilde{A}_\mu^i$ and where the partition functions for the brane sectors are given by:
\begin{align}
&Z_c[\tilde{A}_\mu^i]=\sum_{\left\{\tilde{L}_i\right\}}\delta[\partial_\mu\tilde{j}^\mu_i]
\exp\left\{i\int d^4x\left[-\frac{1}{2m^2}\tilde{j}_{\mu i}^2+\tilde{j}_\mu^i\tilde{A}^\mu_i\right]
\right\},
\label{eq:44}
\end{align}
(without sum in $i$) where the functional $\delta$-distribution enforces the closeness of the monopole worldlines (the chromomagnetic loops) giving the current conservation laws $\partial_\mu \tilde{j}^\mu_i=0$, which are identically satisfied due to (\ref{eq:36}).

An observation is in order at this point. Notice that the activation term for the chromomagnetic loops is highly singular.
This singularity is associated to the hydrodynamic limit, where we consider the coherence length of the condensate to be zero. As discussed in \cite{mvf}, this activation term can be regularized be smoothing out the $\delta$-distributions over the real coherence length of the chromomagnetic condensate, that is non-zero (this, in fact, gives the thickness of the confining chromoelectric flux tube, when external charges are present in this medium), such that the regularized activation term gives an energy contribution proportional to the total length of the chromomagnetic loops $\tilde{L}_i$. Such a regularization can also be done in the explicit evaluation of the confining potential through the introduction of an ultraviolet cutoff scale corresponding to the inverse of the coherence length of the condensate \cite{artigao}.

The third step in our approach consists in the use of the Generalized Poisson's Identity (GPI) (see appendix A of \cite{dafdc} for a detailed discussion on the subject) in $d=4$:
\begin{equation}
\sum_{\left\{\tilde{L}_i\right\}}\delta[\eta_\mu^i-\delta_\mu(x;\tilde{L}_i)]=\sum_{\left\{\tilde{V}_i\right\}}e^{2\pi i\int d^4x\,\tilde{\delta}_\mu(x;\tilde{V}_i)\eta^\mu_i},
\label{eq:45}
\end{equation}
(without sum in $i$) where $\tilde{V}_i$ is the 3-brane Poisson-dual to the 1-brane $\tilde{L}_i$. The GPI works as a brane analogue of the Fourier transform: when the line configurations $\tilde{L}_i$ in the left-hand side of (\ref{eq:45}) proliferate, the volume configurations $\tilde{V}_i$ in the right-hand side become diluted and vice-versa. Using (\ref{eq:45}) we can rewrite (\ref{eq:44}) as:
\begin{align}
&Z_c[\tilde{A}_\mu^i]=\int\mathcal{D}\eta_\mu^i\,\sum_{\left\{\tilde{L}_i\right\}}
\delta\left[\tilde{g}\left(\frac{\eta_\mu^i}{\tilde{g}}-\delta_\mu(x;\tilde{L}_i)\right)\right]\nonumber\\
&\delta\left[\tilde{g}\left(\partial_\mu\frac{\eta_\mu^i}{\tilde{g}}\right)\right]\exp\left\{
i\int d^4x\left[-\frac{1}{2m^2}\eta_{\mu i}^2+\eta_\mu^i\tilde{A}^\mu_i\right]\right\}\nonumber\\
&=\mathcal{N}\int\mathcal{D}\eta_\mu^i\,\sum_{\left\{\tilde{V}_i\right\}}
e^{2\pi i\int d^4x\,\tilde{\delta}_\mu(x;\tilde{V}_i)\frac{\eta^\mu_i}{\tilde{g}}}\int
\mathcal{D}\tilde{\theta}^i\nonumber\\
& e^{i\int d^4x\,\tilde{\theta}^i\partial_\mu\frac{\eta^\mu_i}{\tilde{g}}}
\exp\left\{i\int d^4x\left[-\frac{1}{2m^2}\eta_{\mu i}^2+\eta_\mu^i\tilde{A}^\mu_i\right]\right\}\nonumber\\
&=\mathcal{N}\mathcal{N}'\sum_{\left\{\tilde{V}_i\right\}}\Phi[\tilde{\theta}_\mu^{V i}]\int\mathcal{D}
\tilde{\theta}^i\,\int\mathcal{D}\eta_\mu^i\,\exp\left\{i\int d^4x\right.\nonumber\\
&\left.\left[-\frac{1}{2m^2}\eta_{\mu i}^2-\frac{\eta^\mu_i}{\tilde{g}}
(\partial_\mu\tilde{\theta}^i-\tilde{\theta}_\mu^{V i}-\tilde{g}\tilde{A}_\mu^i)\right]\right\},
\label{eq:46}
\end{align}
(without sum in $i$) where we defined the Poisson-dual current (to the chromomagnetic current $\tilde{j}_\mu^i$), $\tilde{\theta}_\mu^{V i}:=2\pi\tilde{\delta}_\mu(x;\tilde{V}_i)$, being $\mathcal{N}$ a constant associated to the use of the functional generalization of the identity $\delta(ax)=\delta(x)/|a|$ and $\mathcal{N}'$ a constant associated to the fact that there is an overcounting of physically equivalent configurations in the partition function for the brane sector without the brane fixing functional $\Phi[\tilde{\theta}_\mu^{V i}]$ (see \cite{mvf} for a discussion on the subject). Since the constant product $\mathcal{N}\mathcal{N}'$ is canceled out in the calculation of correlation functions and VEV's, we shall effectively neglect them from now on (the partition function is only defined up to global constant factors).

Notice from the geometric interpretation of the GPI given above that the proliferation (dilution) of the ensemble of chromomagnetic worldlines $\left\{\tilde{L}_i\right\}$ is associated to the dilution (proliferation) of the Poisson-dual ensemble of volumes $\left\{\tilde{V}_i\right\}$, what tells us that the Poisson-dual currents $\tilde{\theta}_\mu^{V i}$ must be interpreted as closed chromoelectric vortices describing regions of the spacetime where the chromomagnetic condensate has not been established \cite{mvf,artigao}.

Integrating out the auxiliary fields $\eta_\mu^i$ in the partial partition functions (\ref{eq:46}) and substituting the result back in the complete partition function (\ref{eq:43}) we obtain, as the low energy effective theory for the chromomagnetic condensed regime in the dual picture, the hydrodynamic (or London) limit of a $U(1)\times U(1)$ dual Abelian Higgs model (DAHM):
\begin{align}
Z_c&=\prod_{i=1}^{3}\sum_{\left\{\tilde{V}_i\right\}}\Phi[\tilde{\theta}_\mu^{V i}]\int\mathcal{D}\tilde{B}_\mu^i
\exp\left\{i\int d^4x\right.\nonumber\\
&\left.\left[-\frac{1}{6}(\tilde{G}_{\mu\nu}^i-\Lambda_{\mu\nu}^i)^2+\frac{m^2}{2}\left(\tilde{B}_\mu^i+
\frac{1}{\tilde{g}}\tilde{\theta}_\mu^{V i}\right)^2\right]\right\}\nonumber\\
&=\prod_{i=1}^{3}\sum_{\left\{\tilde{V}_i\right\}}\Phi[\tilde{\theta}_\mu^{V i}]\int\mathcal{D}\tilde{B}_\mu^i
\exp\left\{i\int d^4x\right.\nonumber\\
&\left.\left[-\frac{1}{6}(\tilde{G}_{\mu\nu}^i-L_{\mu\nu}^i)^2+\frac{m^2}{2}\tilde{B}_{\mu i}^2\right]\right\},
\label{eq:47}
\end{align}
(without sum in $i$) where in the last line we made the shift $\tilde{B}_\mu^i:=\tilde{A}_\mu^i-\frac{1}{\tilde{g}} \partial_\mu\tilde{\theta}^i\mapsto\tilde{B}_\mu^i-\frac{1}{\tilde{g}}\tilde{\theta}_\mu^{V i}$. Notice that $m$ is the mass acquired by the gauge invariant field $\tilde{B}_\mu^i$ due to the chromomagnetic condensate. Also, $\tilde{G}_{\mu\nu}^i:=\partial_\mu\tilde{B}_\nu^i-\partial_\nu\tilde{B}_\mu^i$ is the strength tensor field and $L_{\mu\nu}^i:=\Lambda_{\mu\nu}^i+\frac{1}{\tilde{g}}(\partial_\mu\tilde{\theta}_\nu^{V i} - \partial_\nu\tilde{\theta}_\mu^{V i})$ is the so-called chromoelectric brane invariant \cite{jt-cho-su2,artigao}: as discussed before, the chromoeletric brane symmetry corresponds to the freedom of moving the unphysical chromoelectric Dirac branes through the geometric place of the spacetime not occupied by the chromomagnetic monopoles. But since in the chromomagnetic condensed regime the only place not occupied by the chromomagnetic monopoles is the interior of the closed chromoelectric vortices, the chromoelectric Dirac strings are necessarily placed over the closed vortices connected to a pair of probe quark-antiquark. In such a setup, which can be read off from the expression for $L_{\mu\nu}^i$ with non-trivial $\Lambda_{\mu\nu}^i$ (in regions of the spacetime where $\Lambda_{\mu\nu}^i=0$ and $\tilde{\theta}_\mu^{V i}\neq 0$, we have from the expression for $L_{\mu\nu}^i$ the closed chromoelectric vortices), the flux inside the chromoelectric Dirac strings is canceled out by part of the flux inside the closed vortices, leaving as result only open chromoelectric vortices with a pair of probe quark-antiquark in their ends. These open vortices correspond to the confining chromoelectric flux tubes \cite{artigao}.


Notice from (\ref{eq:47}) that it is impossible to realize a complete chromomagnetic condensation (meaning that the monopoles proliferate in such a way that they occupy the whole space) when there are external chromoelectric sources embedded into the system: such a complete chromomagnetic condensation would imply in the complete dilution of the ensemble of internal defects $\left\{\tilde{V}_i\right\}$, what would destroy the brane invariants $L_{\mu\nu}^i$ and spoil the local chromoelectric Dirac brane symmetry and the Elitzur's theorem by ``making the unphysical chromoelectric Dirac strings become real, constituting the confining chromoelectric flux tubes'', what is clearly an absurd. This restriction over the realization of the chromomagnetic condensation in the system in the presence of external chromoelectric sources is easily comprehensible in physical terms: the dual Meissner effect, generated by the mass $m$ of the gauge invariant field $\tilde{B}_\mu^i$ in the condensed regime, expels the chromoelectric fields generated by the external charges of almost the whole space constituted by the dual superconductor, however, these fields cannot simply vanish - they become confined in regions of the space with minimal volume corresponding to the chromoelectric confining flux tubes described by the brane invariants $L_{\mu\nu}^i$ \cite{jt-cho-su2,artigao}.

Before returning to the direct picture, we can rewrite the effective action present in the partition function (\ref{eq:47}) in the Cartan representation as:
\begin{align}
S=\int d^4x \sum_{(k)=3,8} \left[-\frac{1}{4}(\tilde{G}_{\mu\nu}^{(k)}-L_{\mu\nu}^{(k)})^2 + \frac{\tilde{m}^2}{2}\tilde{B}_{\mu(k)}^2\right],
\label{eq:48}
\end{align}
where we defined $\tilde{m}:=\sqrt{\frac{3}{2}}m$. Its dual action is given by \cite{dafdc}:
\begin{align}
*S&=\int d^4x \sum_{(k)=3,8}\left[-\frac{1}{2}(\partial_\mu\tilde{Y}^{\mu\nu}_{(k)})^2
+\frac{\tilde{m}^2}{4}\tilde{Y}_{\mu\nu{(k)}}^2+\right.\nonumber\\
&\left.+\frac{\tilde{m}}{2}\tilde{Y}_{\mu\nu}^{(k)}L_{(k)}^{\mu\nu}\right],
\label{eq:49}
\end{align}
where the Kalb-Ramond fields $\tilde{Y}_{\mu\nu}^{(k)}$ describe the monopole condensate in the direct picture: notice the \emph{rank-jumping} observed in the direct picture due to the monopole condensation - in the diluted regime described by (\ref{eq:29}) the system is characterized by the massless 1-forms $A_\mu^{(k)}$, while in the condensed regime described by (\ref{eq:49}) the system is characterized by the massive 2-forms $\tilde{Y}_{\mu\nu}^{(k)}$. The rank-jumping is a signature of the defects condensation and the mass generation in the JTA \cite{mcsmon,dafdc,jt-cho-su2,artigao}. Notice also that (\ref{eq:49}) is the generalization of \cite{qt} compatible with the Elitzur's theorem and the local chromoelectric brane symmetry: in \cite{qt} the last term in (\ref{eq:49}) features a minimal coupling of the Kalb-Ramond field directly with the chromoelectric Dirac strings instead of the chromoelectric brane invariants, thus violating the Elitzur's theorem and the local chromoelectric brane symmetry by ``making the unphysical chromoelectric Dirac strings become real, constituting the confining chromoelectric flux tubes''.

The partition function describing the chromomagnetic condensed regime in the direct picture is, then, given by:
\begin{align}
Z_c&=\prod_{(k)=3,8}\sum_{\left\{\tilde{V}_{(k)}\right\}}\Phi[\tilde{\theta}_\mu^{V (k)}]\int\mathcal{D}\tilde{Y}_{\mu\nu}^{(k)}
\exp\left\{i\int d^4x\right.\nonumber\\
&\left.\left[-\frac{1}{2}(\partial_\mu\tilde{Y}^{\mu\nu}_{(k)})^2
+\frac{\tilde{m}^2}{4}\tilde{Y}_{\mu\nu{(k)}}^2+\frac{\tilde{m}}{2}\tilde{Y}_{\mu\nu}^{(k)}L_{(k)}^{\mu\nu}\right]\right\},
\label{eq:50}
\end{align}
(without sum in $(k)$). As discussed before, it is impossible to realize a complete chromomagnetic condensation in the presence of external chromoelectric sources: the best the dual Meissner effect can do is to completely dilute the closed chromoelectric vortices disconnected from the chromoelectric Dirac strings. In such a case, by integrating out the fields $\tilde{Y}_{\mu\nu}^{(k)}$ in (\ref{eq:50}), we obtain \cite{artigao}:
\begin{align}
Z_c&=\prod_{(k)=3,8}\exp\left\{i\int d^4x\left[-\frac{1}{2}j^{(k)}_\mu\frac{1}{\partial^2+\tilde{m}^2}j_{(k)}^\mu\right]\right\} \nonumber\\
&\sum_{\left\{\bar{S}_{(k)}\right\}}\exp\left\{i\int d^4x \left[-\frac{\tilde{m}^2}{4}L^{(k)}_{\mu\nu}\frac{1}{\partial^2+\tilde{m}^2}L_{(k)}^{\mu\nu}\right]\right\},
\label{eq:51}
\end{align}
(without sum in $(k)$) where the geometric sum is now taken over all the possible shapes of the confining flux tubes. This sum is very difficult to realize in general. However, if we consider a static probe quark-antiquark configuration with a spacial separation $L$ and the asymptotic time regime $T\rightarrow\infty$, then it is reasonable to approximate the sum over brane invariants in (\ref{eq:51}) by taking into account only its dominant contribution, which is given by a linear flux tube of lenght $L$ corresponding to the stable asymptotic configuration which minimizes the energy of the system \cite{artigao}. In this limit, one obtains from (\ref{eq:51}) the following static interquarks potential (see \cite{artigao} for the detailed evaluation):
\begin{align}
V_{static}(L;g,\tilde{m},\tilde{M})=\sigma(g,\tilde{m},\tilde{M})L-
\frac{(Q_{(3)}^2+Q_{(8)}^2)}{4\pi}\frac{e^{-\tilde{m}L}}{L},
\label{eq:52}
\end{align}
where the string tension is given by:
\begin{align}
\sigma(g,\tilde{m},\tilde{M})&=\frac{(Q_{(3)}^2+Q_{(8)}^2)\tilde{m}^2}{8\pi}\ln\left(\frac{\tilde{m}^2+\tilde{M}^2}
{\tilde{m}^2}\right)\nonumber\\
&=\frac{g^2\tilde{m}^2}{24\pi}\ln\left(\frac{\tilde{m}^2+\tilde{M}^2}{\tilde{m}^2}\right),
\label{eq:53}
\end{align}
where $\tilde{M}$ is an ultraviolet cutoff corresponding to the Higgs mass (the mass of the monopoles), whose inverse gives the coherence length of the monopole condensate and where we used the fact that for any mesonic configuration ($r-\bar{r}$, $b-\bar{b}$ or $y-\bar{y}$), we have $Q_{(3)}^2+Q_{(8)}^2=\frac{g^2}{3}$.

The static interquarks potential (\ref{eq:52}) was originally obtained in \cite{suganuma}, where the free parameters $(g,\tilde{m},\tilde{M})$ were fixed by fitting the profile of the phenomenological Cornell potential as being $(5.5,0.5 GeV,1.26 GeV)$. This set of values reproduces the experimental value of the string tension, $\sigma\approx (440 MeV)^2$, obtained from the slope of the Regge trajectories, and gives the prediction that the dual superconductor realizing the static chromoelectric confinement in the QCD vacuum should be of the type II. Notice also that taking $\tilde{m}=0$ leads us back to the diluted regime characterized by the long-range Coulomb interaction, eliminating the monopole condensate and destroying the chromoelectric confinement.

It is important to stress here the two main differences between our approach and the approach of \cite{suganuma}:

a) Our work has as the starting point the Abelian action with chromomagnetic monopoles (\ref{eq:35}), obtained using the $SU(3)$ Cho decomposition and the discarding of the off-diagonal sector of the theory parametrized by the valence potential, while the starting point in \cite{suganuma} is the Abelian action with chromomagnetic monopoles obtained in \cite{suzuki-su3} using the $SU(3)$ Abelian projection implemented specifically in the MAG. Since the Abelian projection method involves a partial gauge fixing condition, there is an ambiguity involved in the choice of a particular Abelian gauge and in the corresponding definition of the monopoles as discussed in \cite{chernodub}: the different Abelian gauges that can be fixed in the Abelian projection method lead, in general, to different results for the string tension. In particular, the Abelian string tension obtained in the MAG in the $SU(2)$ case gives support to the Abelian dominance hypothesis. However, this result is obscure from a physical point of view, since the value of a physical observable like the string tension should not depend on an arbitrary gauge choice. On the other hand, the Cho decomposition allows one to reveal the monopoles in the non-Abelian theory without resorting to any gauge fixing procedure, what represents an apparent advantage over the usual Abelian projection method, since we do not have an ambiguity in the choice of a particular Abelian gauge and in the definition of the monopoles via the Cho decomposition. In fact, if the off-diagonal sector of the theory can be discarded at all in some regime of the theory, then the result obtained for the string tension using the Cho decomposition should be unique in principle, corresponding to the result found in our work, which agrees with the result of the Abelian projection method implemented specifically in the MAG. Although the effective potential is the same in both cases, via the Abelian projection method the choice of the MAG is in principle only one between many different possibilities, while via Cho decomposition the result we have obtained is in principle unique. In this sense, we see our result as being gauge independent;

b) In what concerns specifically to the evaluation of the confining potential, there is another important conceptual difference between the procedure of \cite{suganuma} and ours. In the calculation of \cite{suganuma}, the chromoelectric Dirac string is taken as being the linear confining chromoelectric flux tube. This cannot be correct as a matter of principle, since the Dirac string is non-physical. As discussed here, the confining flux tubes correspond actually to open chromoelectric vortices with a pair of probe quark-antiquark in their ends, which are described in our formalism by brane invariants. These open vortices emerge in our formalism due to the mutual cancellation between part of the chromoelectric flux inside the closed chromoelectric vortices connected to the Dirac strings and the chromoelectric flux inside the strings. This mutual cancellation necessarily happens due to the Dirac's veto. This is a formal advance featured in our approach. Notice also that in the partition function (\ref{eq:51}), all the possible shapes of the confining flux tubes contribute in the sum over the brane invariants. Hence, in general, our result is different from the result of \cite{suganuma}, which takes into account only the straight shape for the string. However, for a static probe quark-antiquark configuration in the asymptotic time regime $T\rightarrow\infty$, as discussed in details in \cite{artigao}, the dominant contribution in the sum over configurations of the brane invariants is given by the tube with the minimal volume (stable configuration that minimizes the energy of the system), which corresponds to a straight flux tube. In this limit, it is reasonable to approximate the effective static interquarks potential by taking into account only the contribution of the straight tube, which is the basic consideration that makes the form of our effective potential equivalent to the one obtained in \cite{suganuma}.

\section{Concluding discussion}
\label{sec:conclusion}

In this Letter we used a generalization of the Julia-Toulouse approach for condensation of defects to study how the confinement of static external chromoelectric probe charges could emerge from the Abelian sector of the pure $SU(3)$ gauge theory due to the condensation of chromomagnetic monopoles.

We took as the starting point to the novel approach here presented, regarding the monopole condensation, the expression for the restricted $SU(3)$ gauge theory defined by means of the Cho decomposition of the non-Abelian connection. This decomposition consists in a reparameterization of the non-Abelian connection which reveals its Abelian sector and the associated topological structures (monopoles) without resorting to any gauge fixing procedure, hence providing a gauge invariant definition of these defects in the Yang-Mills theory.

With the discarding of the off-diagonal sector of the theory (assuming the validity of the hypothesis of Abelian dominance), we showed that the action in the regime with diluted defects can be put in the form of a Maxwellian theory non-minimally coupled to chromomagnetic monopoles and minimally coupled to external chromoelectric probe charges. This was the crucial point that allowed us to apply the generalized form of the Julia-Toulouse approach for condensation of defects and obtain a hydrodynamic effective theory for the regime where the monopoles are condensed. The effective theory that we derived with such an approach gives an interaction potential between two static chromoelectric probe charges of opposite signs embedded in the chromomagnetic monopole condensate consisting in a sum of a Yukawa and a linear confining term in the asymptotic time regime $T\rightarrow\infty$. Our result is in principle gauge independent.

\section{Acknowledgements}

We thank L. E. Oxman for helpful discussions about the Cho decomposition and Conselho Nacional de Desenvolvimento Cient\'ifico e Tecnol\'ogico (CNPq) for financial support.

\end{document}